# An Anti-attack Model Based on Complex Network Theory in P2P networks

Hao Peng[a], Songnian Lu[a], Dandan Zhao[a], Aixin Zhang[b], Jianhua Li[a,b]

[a] *Department of Electronic Engineering，Shanghai Jiao Tong University, Shanghai, China*

[b] *School of Information Security Engineering，Shanghai Jiao Tong University, Shanghai, China*

**Abstract:** Complex network theory is a useful way to study many real systems. In this paper, an anti-attack model based on complex network theory is introduced. The mechanism of this model is based on dynamic compensation process and reverse percolation process in P2P networks. The main purpose of the paper is: (i) a dynamic compensation process can turn an attacked P2P network into a power-law (PL) network with exponential cutoff; (ii) a local healing process can restore the maximum degree of peers in an attacked P2P network to a normal level; (iii) a restoring process based on reverse percolation theory connects the fragmentary peers of an attacked P2P network together into a giant connected component. In this way, the model based on complex network theory can be effectively utilized for anti-attack and protection purposes in P2P networks.

## 1 Introduction

In recent years, Peer-to-Peer (P2P) networks become a very active research area due to its advantages over traditional Client/Server model for application scenarios such as file sharing, distributed computing, collaborative applications, and immediate communication [1-6]. However, they are vulnerable to many malicious peers for the open and unreliable nature of most P2P networks. For example, in P2P file sharing systems, malicious peers can spread untrustworthy and potentially harmful files to the system if no security measure is used [1]. And the same situation happens in distributed computing systems [2], an attacked computer can affect many other computers' normal operations. For these reasons, during the process of gaining service from P2P networks, how to resist attacks in the study of P2P networks becomes more and more important.

In order to avoid being attacked in P2P networks, there are mainly two kinds of anti-attack models: trust-based an-attack model [7-8] and anonymity-based an-attack model [9-10]. How to make good use of peers' identity privacy became an important research basis and the designing mechanism of the both models relates to peers' identity information. In this way we inevitably increase the network's computational overhead and thus they are both difficult to achieve in large complex P2P networks such as fully distributed P2P networks [11]. Therefore a peer's identity information can't be obtained by other peers for considering safety factors. Considering these factors above, we need to find other ways to achieve our security and distributed goal such as complex network theory [12].

In this paper, we proposed an anti-attack model based on complex network theory in P2P networks: (i) we analyze the characteristics of attack in P2P networks. According to our analysis, the type in this kind of networks is progressive attack; (ii) we use the PL (Power Low) character of complex network to design a healing mechanism. We shift the focus to the study of progressive and rapid attacks on networks that can respond in an active fashion. In the process, we design several techniques that managed networks can adopt to minimize the damages, and also to efficiently recover from the aftermath of successful attacks. We find that the PL nature of the networks can be judiciously utilized for such protection and recovery purposes. Our simulations show this mechanism can recover the PL exponent of P2P network. After that, using generating functions and reverse percolation, we restore connectivity of P2P network.

The outline of the paper is as follows. In Sec. 2 we introduce the anti-attack model based on complex networks. In Sec. 3 we show how percolation ideas and generating function methods can be used to provide exact solutions of these models on simple networks with uncorrelated transmission probabilities. In Sec. 4 we give our discussion and conclusions.

## 2 Proposed Model Based on Complex Network Theory

## 2.1 Notions

The main notations used in this paper are listed according to their orders in this paper as follows.

$\alpha$ —The exponent of a destroyed P2P network before a healing process.

$\beta$ —A stretch factor of the maximum degree in a healing process.

$d_i$—The degree of the peer $p_i$.

$P(d), P^{'}(d)$ —The degree distribution of a P2P network before and after the healing process.

$d_0$, $d_{max}$—The maximum of a destroyed P2P network before and after a healing process.

$M(k)$—The size distribution of some small components in a restored P2P network.

$N(k)$—The degree distribution of the process of a reverse percolation.

## 2.2 Basic Model Design

This section we consider a P2P network where the set of high degree peers is removed after being attacked. Then we show how a local healing process can restore the maximum degree to an existing cutoff. Each peer decides randomly and independently how many preferential edges to insert and in this case the healing process doesn't need any central coordination mechanism.

We assume a destroyed P2P network' exponent be $a$ and its maximum degree $d_0$. The purpose of the healing process is to increase this maximum degree of the distribution by a stretch-factor $\beta = d_{max}/d_0$. Thus the maximum degree after the healing process will increase from $d_0$ to $d_{max}$ while the exponent remains the same. Here each peer $p_i$ independently decides to compensate with probability $f$. Next the healing process involves making $(\beta-1)d_i$ new preferential connections where $d_i$ is the degree of the peer $p_i$. Let $P(d)$ and $P'(d)$ be the degree distributions before and after the healing process respectively. In this way we can see that given $P(d)$ is a P2P network with maximum degree $d_0$ and $f = \beta^{-a}$ and then $P'(d)$ follows a power-low distribution as $\lambda(a,\beta)d^{-a}$ where $\lambda$ is a constant depending on $a$ and $\beta$. Then if the degree distribution is power-law to begin with, this healing process will not change the distribution.

Thus, while the attack bounds the maximum degree of the distribution, the healing scheme is able to preserve the exponent of the power-low distribution of the P2P network. As long as each incoming peer makes $n > 1$ random preferential connections, there must exists a giant connected component even for very high rates of preferential attacks (see Table 1); thus the loss of connectivity is not one of the damaging effects of such attacks. The healing scheme requires two steps as follows:

Step 1: A new peer $p'$ is added to the network and makes m preferential attachments.

Step 2: With probability $r$ a preferentially selected peer $o$ is chosen. The preferential selection process selects this peer $o$ with degree $d(o)$ and probability $d(o)/\sum_i d(i)$. The selected peer $o$ is then deleted from the network and the deletion process is: delete peer $o$ and all its edges. Then $o$ starts as a new peer and makes some preferential attachments; for all peers that were connected to $o$ which need to lose an edge and each one compensates by adding an edge preferentially.

We have performed Monte Carlo simulations to validate the results where at each step a peer is added and makes $n = 2$ preferential attachments. Then with probability $r$ we delete a peer and choose a peer to be deleted preferentially.

Table 1: $g(r)$, the scale of peers in the largest connected component for various values of $r$.

| $r$ | 0.02 | 0.05 | 0.08 | 0.11 | 0.2 | 0.3 |
|---|---|---|---|---|---|---|
| $g(r)$ | 1.0 | 0.997 | 0.993 | 0,989 | 0.983 | 0.971 |

From the table 1, we can see that connectivity of P2P networks is preserved even for large values of $r$.

## 2.3 Implementations

In this section，we briefly introduce a scenario to demonstrate how our model works. The scenario we refer to is that we are given a network with a sharp cutoff and the peers randomly decide to introduce new edges to restore the cutoff to a desired value $d_{max}$. Each peer can react whenever it loses an edge unannounced due to an attacked neighbor being deleted. Then we use the feedback scheme to proceed to the heeling process when the peer encounters the purposed attacks.

This process includes the following steps.

(1) Each peer $p$ performs the healing algorithm of Section 2.2 with probability $1/k_i$ when losing an edge, where $k_i$ is the degree of the $i_{th}$ peer.

(2) Neglect the degree correlation of the peers, the probability of any edge being deleted when peers of degree greater than $k_0$ are deleted will be given by:

$$\overline{p} = \frac{\sum_{k_0}^{k_{max}} kP(k)}{E} = \xi k_0^{2-\sigma}$$

Here $\xi$ is a constant in the order of 1. Then the probability of a peer with degree $k$ losing an edge is $P_k \approx k\overline{p}$.

(3) We can compute the probability of a peer beginning with the healing process is approximately $P_k / k = \overline{p}$ which depends on the level of the attack. In order to simply our focus, we don't consider second-order effects and degree correlations here.

As been discussed above, our model is basically designed to detect any large-scale deletion of a fraction of network edges. Therefore, the proposed model ensures that a certain fraction of the peers will always perform the healing algorithm in the case of a large-scale attack.

## 2.4 Simulation

In order to evaluate the effectiveness of the proposed mechanism above, a software simulator built from scratch is adopted. Here we have performed the Monte Carlo simulations of the healing process, which is used to analyze the classic simulation model of complex network. In our simulation design, we use a mesh topology with 50000 peers selected randomly. This mesh represents a general topology and it can also be applied to specific P2P networks[13]. The simulator relies on a discrete time paradigm and the time step is equal to 225 ms.

To perform the simulation analysis, we adopted the following parameter values. For the sake of clarity only ten minutes of the overall simulation is presented. To obtain a realistic simulation we limited the available bandwidth. According to the application characteristics of P2P networks, the bandwidth is unable to keep a sustained speed of 5.00 Mb/s, but rather tends to stabilize around a maximum 2.75 Mb/s. The movement of all peers was randomly generated with a maximum speed of 2.50 Mb/s and an average pause of 30s. Each simulation runs 500 simulation seconds. The result is shown in Fig. 1~Fig. 3. The vertical axis shows the average degree of the component of peers in different value (Log), while the horizontal axis shows the number of peers (Log).

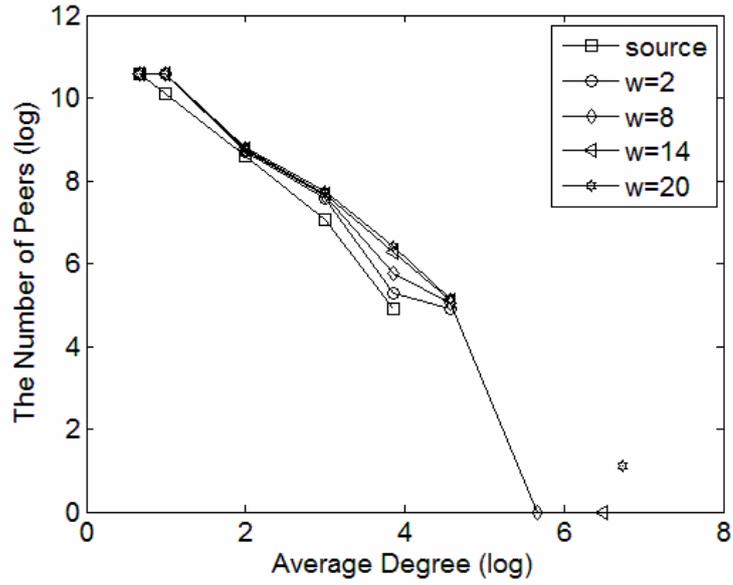

Fig.1 Degree distribution before and after healing process when $a$ =2.4 and the stretch-factors $w$ from top to down is: $w$ =2, 8, 14, 20, respectively

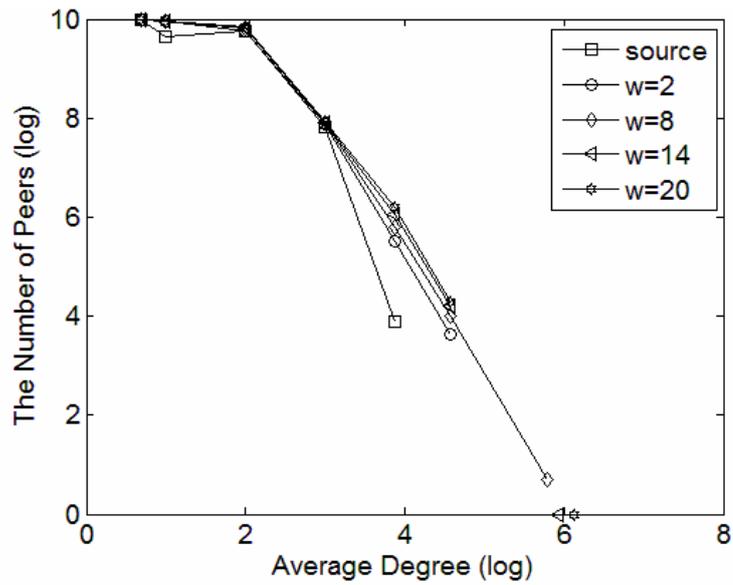

Fig.2 Degree distribution before and after healing process when $a$ =2.8 and the stretch-factors $w$ from top to down is: $w$ =2, 8, 14, 20, respectively

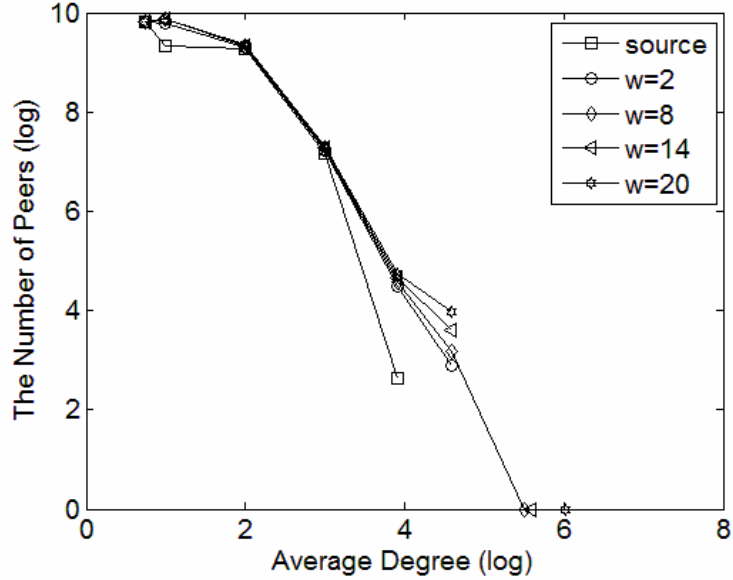

Fig.3 Degree distribution before and after healing process when $a$ =3.2 and the stretch-factors $w$ from top to down is: $w$ =2, 8, 14, 20, respectively

As mentioned above, Fig. 1~Fig. 3 shows that how the healing algorithm works for different values of $a$. The source P2P network is an attacked network where the set of high degree peers is removed after being attacked. In order to restore this kind of network, according to the basic model introduced above, we add some preferential attachments to heel it. Here we set the number $w$ =2, 8, 14, 20 respectively to compare the different heeling effects. Simulation results show that a small number of new edges can result in the power-low character of the attacked network being restored efficiently in Fig. 1.

However, this doesn't mean that our heeling mechanism can work completely with the same number of stretch-factors when a power-low P2P network with greater exponent in Fig.2 and Fig.3. As we know, a power-low P2P network that has a large exponent shows it has some very high degree of peers and if these peers have been attacked, the power-low character of the network must be changed immensely. Therefore, we can't add too many random edges to restore because more preferential attachments can lead to less real reflection of the attacked network when the power-low exponent becomes larger. Then that we set $w$ =14 when $a$ =2.8 in Fig. 2 is a good choice to restore the attacked network, accordingly we set $w$ =8 when $a$ =3.2 in Fig. 3.

## 3  Restoring connectivity in P2P networks

In section 2, we introduce that a power-low P2P network will heavily destroyed when the set of high degree peers is removed after being a targeted attack. In our design above, we insert some new edges randomly and then perform a local heeling process to restore the set of high degree peers. However, another question is how to restore the connectivity when the attacked network has been heeled and how many connections among the peers we need to form a giant connected component in a destroyed P2P network.

Thus our work in this section is to show that how to quick response from a targeted attack to restore the connectivity. Besides, we will indicate that a very small number of new connections are enough to quickly reconnect most of the broken power-low P2P network even under very large-scale targeted attacks. As we know, a proof for an attacked power-low P2P network being attacked is difficult to attain and then we assume that for a simulating power-low P2P network, the different-scale removal of the highest degree peers indeed creates disconnected components with different distribution of the size of components. Finally, we indicate that as long as we have knowledge about the size distribution with heavy-tailed of components of peers and then only a small number of random connections being inserted will connect the fragments of components together into a giant connected component.

## 3.1 The size distribution of components of peers after attack

In general, a targeted attack in P2P networks is hard to follow and analyze where there are no central mechanism in this kind of network and thus some attacked components of peers can't respond to this incident immediately. However, the statistics of the fragmented components of peers after a targeted attack can be followed. In this section, according to complex network theory [14][15], we now show that the size distribution of the connected components in P2P networks on a given degree distribution can be expressed using the generating function forms.

Now we start by examining attack-percolation process for the general case in which occupation probability (the probability of a peer being attacked by some malicious peers) is an arbitrary function of the peer degree. Let $p_k$ be the probability that a randomly chosen peer has degree $k$, and $q_k$ be the probability that a peer is attacked provided that it has degree $k$. Then we can define the generating function forms of a destroyed P2P network after a targeted attack as follows:

$$G_0(x) = \sum_{k=1}^{\infty} q_k p_k x^{-k} = \sum_{k=1}^{d_{max}} q_k p_k x^{-k}$$

$$G_1(x) = \frac{G_0'(x)}{G_0'(1)}$$

Here $d_{max}$ is the maximum degree of the peers in the specific simulation sample of a P2P network. For the family of PL P2P networks, we assume that the cutoff degree of the peers diverges as the maximum degree $d_{max}$ in the P2P network goes to infinity. Therefore the above equations can define the feature of relatively large cutoff degree in a PL P2P networks. In particular, if $q_k$ decreases with $k$, it can define a form of the targeted attack in effect. Then in our design, we let an attack form for which $q_k = mk^{-q}$ where $q$ is an analytical parameter of that how targeted the attack is and $m$ is a stretch constant. Thus the generating functions of an attacked P2P network after attack where an original P2P network with exponent $\theta$ can be simply derived as:

$$G_0(x) = \sum_{k=1}^{d_{max}} m'k^{-q-\theta}x^{-k}$$

$$G_1'(1) = \sum_{k=1}^{d_{max}} m''k^{-q-\theta+2} \propto O(d_{max}^{-q-\theta+3})$$

Where $m'$, $m''$ are both positive constant and $q+\theta > 3$. Especially if the form of an attack is the linear targeted attack, the attack parameter $q=1$ and then we can show that the value above is always finite for $\theta > 2$. It can be seen that that such attack will in fact increase the value of $\theta$ by a certain value $q$. Therefore following the way of Aiello et al. in Ref. [16], we can come to the conclusion that there will be no giant connected component of peers in an attacked P2P network if $q+\theta > \beta_c = 3.4785$. However, using the restoring connectivity process in Section 3.2, we can obtain the size distribution of the connected component of peers after a targeted attack.

From what has been discussed above, for an attacked PL P2P network using the generating function forms as described above, we can obtain the distribution of the fragmented components even if the attacked P2P network might not have any giant connected component of peers and the size distribution of the destroyed P2P network could still be followed. Next we will discuss how to restore the fragmented P2P network to a normal P2P network which has an original giant connected component with the same power low and same topological structure.

## 3.2 The restoring connectivity process

In this section, we will make use of percolation theory [17] as the basis of restoring connectivity of an attacked P2P network. In this way the destroyed network is divided into many small fragments. In the percolation theory, we need to add random edges among many small components of the peers in the P2P network. As the number of such random links increases to a certain threshold, different components of the network must start to bind together until a giant connected component occurs and thus these giant connected components contain most of the peers in the attacked network. In this way the connectivity process can work as we expect and the specific detail can be described as following parts.

Firstly we let $M(k)$ be the size distribution of these small components and it is also the probability of these components that have size $k$. We can conclude that the process corresponds to a percolation on a random graph with degree distribution $M(k) \equiv N(k)$. To obtain a random graph with a prescribed degree distribution $N(k)$, we can design as follows: for any $k$, there will be $nN(k)$ peers of degree $k$ in the network, arranged as $A_1^k, A_2^k, A_3^k ... A_{nP(k)}^k$. Each of these peers creates $k$ dummy duplicate peers where the duplicates of the $i^{th}$ peer are expressed as $A_{1,i}^k, A_{2,i}^k, A_{3,i}^k,...,A_{k,i}^k$. Then each peer of dummy copies can start a random matching while all the dummy peers correspond to a real peer. After the matching, it will collapse the duplicates of a peer into its real peer and thus all the links to the duplicates will now be links to the real peer itself.

From what has been discusses above, the process of the restoring connectivity in P2P networks can be readily achieved. In the process, any reconnected component can be viewed as the duplicate peers of a real peer and the insertion of random edges can be showed as a percolation process whose degree distribution is equal to the distribution of the component sizes of the real broken components. In the usual percolation process, keep each edge with probability $p$ which in the random graph construction process, basically involves doing random matching with probability $p$ in the network with duplicated peers. With this correspondence, many of the well-known results for the percolation process on random graphs with a given degree distribution can be readily applied to the restoring process.

In our design, the percolation threshold for the probability of successful attempts required for a giant component to appear can be calculated as: $q_c = <C>/(<C^2> - <C>^2)$ where $<C>$, $<C^2>$ are the average and variance of the distribution of the connected components[8]. While the average size of the connected components in an attacked P2P network is finite and therefore it is possible for the variance of this distribution to be very large. In this way, the corresponding percolation threshold will be small. In particular, if the distribution of the size of the connected components follows a power-low distribution with exponent $2.4 < \theta < 3.7$ and the maximum component size $C_{max} \gg 1.2$. Thus we are sure that most of the disconnected components are reconnected and the connectivity of the broken network can be restored.

## 4  Discussion and Conclusions

In this paper, we proposed an anti-attack model based on complex network theory in P2P networks. It can restore the maximum degree and topological structure in a local healing manner, and restore the connectivity after a P2P network being attacked. Therefore our research provides one way on the dealing with attacks in P2P networks and it can ensure P2P network's communication security and robustness.

However, the mechanism proposed in the paper only considering a single kind of P2P networks that does not interact and depends on other kinds of P2P networks to avoid attackers. In reality, there are many interdependent P2P networks interacting with each other. To enhance the our design here, in future work, we will introduce other mechanisms such as the robustness of different kinds of networks [18] for studying interacting P2P networks and will present an exact anti-attack model for a single P2P network of n interdependent P2P networks. In this way, our model can be optimized.

# References

[1] S. Lee, S.-H. Yook, Y. Kim, Physica A 385 (2007) 743.

[2] S. Blanas, V. Samoladas, Further Generation Computer Systems 25 (2009) 100.

[3] J.X.Gao, Z.Chen, Y.Z.Cai, X.M.Xu, Phys.Rev.E 81(2010)041918.

[4] R. Gaeta, M. Sereno, Concurrency and Computation-Practice & Experience 20 (2008) 713.

[5] E. Pournaras, G. Exarchakos, N. Antonopoulos, Computer Communications 31 (2008) 3030.

[6] F. Wang, Y.R. Sun, Computational Intelligence 24 (2008) 213.

[7] E. Bertino, E. Ferrari, A. Squicciarini, IEEE Transactions on Knowledge and Data Engineering 16(2004) 827.

[8] S.S. Song, K. Hwang, R.F. Zhou, IEEE Internet Computing 9(2005) 24.

[9] CY Chow, MF Mokbel, XA Liu, Geoinformatica 15(2011) 351.

[10] J. Zhang, H.X. Duan, W. Liu, Computer Communications 34(2011) 358.

[11] T. Enokido, A. Aikebaier, M. Takizawa, IEEE Transactions on Industrial Electronics 58(2011)2097.

[12] X.Huang,J.X.Gao,S.V.Buldyrev,S.Havlin,H.E.Stanley,Phys.Rev.E 83(2011)065101.

[13] E.K. Lua, J. Crowcroft , et al., IEEE Communications on Survey and Tutorial 7(2005)72.

[14] M.E.J. Newman, S.H. Strogatz, D.J. Watts, Phys. Rev. E 64 (2001) 026118.

[15] S.H.S.D.S. Callaway, M.E.J. Newman, D.J. Watts, Phys. Rev. Lett. 85 (2000) 5468.

[16] W. Aiello, F. Chung, L. Lu, The 32nd Annual ACM Symposium on Theory of Computing, ACM (2000) 171.

[17] M. Molloy, B. Reed, Random Struct. Algorithms 6 (1995) 161.

[18] J.X.Gao,S.V.Buldyrev,S.Havlin,H.E.Stanley, arXiv: 1010.5829:(2010).